\documentclass[fleqn, 10pt]{wlscirep}

\usepackage{amsmath}
\usepackage{amssymb}
\usepackage{graphicx}
\usepackage{color}
\usepackage{bm}
\usepackage{soul}
\usepackage{color}

\title{Memory Remains: Understanding Collective Memory in the Digital Age}

\author[1,+]{Ruth  Garc\'{i}a-Gavilanes}
\author[2,+]{Anders Mollgaard}
\author[1]{Milena Tsvetkova}
\author[1,3,*]{Taha Yasseri}
\affil[1]{Oxford Internet Institute, University of Oxford, Oxford, UK}
\affil[2]{Niels Bohr Institute, Copenhagen, Denmark}
\affil[3]{Alan Turing Institute, London, UK}
\affil[*]{Corresponding Author: taha.yasseri@oii.ox.ac.uk}
\affil[+]{these authors contributed equally to this work}

\begin{abstract} 
Recently developed information communication technologies, particularly the Internet, have affected how we, both as individuals and as a society, create, store, and recall information. Internet also provides us with a great opportunity to study memory using transactional large scale data, in a quantitative framework similar to the practice in statistical physics. 
In this project, we make use of online data by analysing viewership statistics of Wikipedia articles on aircraft crashes. We study the relation between recent events and past events and particularly focus on understanding memory triggering patterns. We devise a quantitative model that explains the flow of viewership from a current event to past events based on similarity in time, geography, topic, and the hyperlink structure of Wikipedia articles. We show that on average the secondary flow of attention to past events generated by such remembering processes is larger than the primary attention flow to the current event. We are the first to report these cascading effects. 
\end{abstract}
\begin{document}
\flushbottom
\maketitle 
\thispagestyle{empty}


\section*{Introduction}
Memory and the way individuals remember, forget, and recall events, people, places, etc. have been prominent topics of theoretical research for a long time \cite{ebbinghaus1885gedachtnis}. However, the notion of collective memory as a socially generated common perception of an event has been introduced and studied only recently \cite{halbwachs1992collective}, about the time when our societies started to become highly connected through new channels of communication. Most of previous studies concern offline settings. However, developments in digital technologies in recent years have significantly influenced how we keep track of events both as individuals and as a collective. Digital technologies have also provided us with huge amounts of data, which researchers are already using to study different aspects of our social behaviour. 

The Internet doesn't forget. On the one hand, the Internet has had strong impacts on memory and the processes of remembering and forgetting and on the other, it has converted collective memory into an observable phenomenon that can be tracked and measured online at scale. Analysing different Web documents, researchers have shown that more recent past events are remembered more vividly in the present. For example, \cite{Yeung2011,Jatowt:2013} investigated news corpora and concluded that most of the temporal expressions are from the near past. The authors of ~\cite{Campos11whatis} analysed ~63K web query logs and found that 10\% had temporal references, mostly to the near past or future. Further, in \cite{Jatowt:2015} the authors studied how microbloggers collectively refer to time and found that although several posts are about past events, the ``here and now" is what they mostly refer to and care about.

Aiming to enhance our knowledge of online collective memory, we use page view logs of articles on Wikipedia, the largest online encyclopaedia.These data provide remarkable granularity and accuracy to study online memory. There is a high correlation between search volume on Google and visits to Wikipedia articles related to the search keywords \cite{Ratkiewicz2010,Yoshida2015}. This indicates that Wikipedia traffic data reliably reflect web users’ behaviour in general. The high response rate and pace of coverage in Wikipedia in relation to breaking news \cite{Althoff2013,KeeganHow2011} is another feature that makes Wikipedia a good research platform to address questions related to collective memory. 

Indeed, other researchers have previously used Wikipedia to study collective memory. In particular, Ferron~\emph{et al.}~\cite{Ferron2011,Ferron2011b, Ferron2012} studied thoroughly editors' behavior to confirm the interpretation of Wikipedia as a global memory place. They explored edit
activity patterns with regard to commemoration processes, the sentiments of edits in old and recent traumatic and non-traumatic events, and the evolution of emotions in talk pages. This work, however, focuses on editorial activities on Wikipedia; only few studies address collective memory considering Wikipedia visitors and their patterns of attention. For example, Yucesoy and Barab{\'{a}}si~\cite{Yucesoy2016} used Wikipedia viewership data to study the popularity and fame of current and retired elite athletes and found that performance dictates visibility and memory. 
More specifically, Kanhabua \emph{et al.}~\cite{Kanhabua2014} tackle remembering signals using page views in Wikipedia to identify factors for memory triggering. They calculate a remembering score made up of different combinations of time series analysis techniques and study how that score varies with regard to time and location. However, this work stops at the empirical observations and fails to give any general understanding of the phenomenon. 

Several other prediction tasks have been done using Wikipedia data and metadata. For example, researchers have used Wikipedia viewership data to predict movie box office revenues~\cite{MartonEarly2013}, stock market moves \cite{moat2013}, electoral popularity~\cite{YasseriB14,YasseriB16}, and influenza outbreaks~\cite{McIver2014,hickmann2015forecasting}.
Further, researchers have predicted the click-through rate between Wikipedia pages, which allows to point which existing and potential Wikipedia links are useful. They have done this using web server logs~\cite{ Paranjape2016} and navigational paths~\cite{west2015mining}. Researchers have also used page view counts to predict the dynamics of Wikipedia pages. For example, \cite{Thij2012} predicted that the attention to promoted content on Wikipedia decays exponentially over time. 

 Using Wikipedia viewership data, we study how new events trigger a flow of attention to past events, which is how we operationalise collective memory. We limit our focus to aircraft incidents and accidents as reported in English Wikipedia, which is the largest language edition of the online encyclopaedia. We quantify and model the attention that flows from articles about recent accidents to articles about past accidents and study the effect of different dimensions of the event on the distribution of attention flow.


\section*{Results}
\label{sec:results}

To calculate the effect of a new event on the attention to a past event, we pair the page view time series of the corresponding Wikipedia articles. Here, we focus on all 
aircraft incidents or accidents reported in English Wikipedia. We call  the events 
that occurred within 2008-2016 ``source events" and their Wikipedia articles ``source articles." We pair the source events with older aircraft incidents or accidents, called ``target events", and their Wikipedia articles, or ``target articles" (see Data and Methods).

\subsection*{View Flow}
\label{subsec:viewflow}

\begin{figure}[ht]
	\centering
	\includegraphics[width=0.95\textwidth]{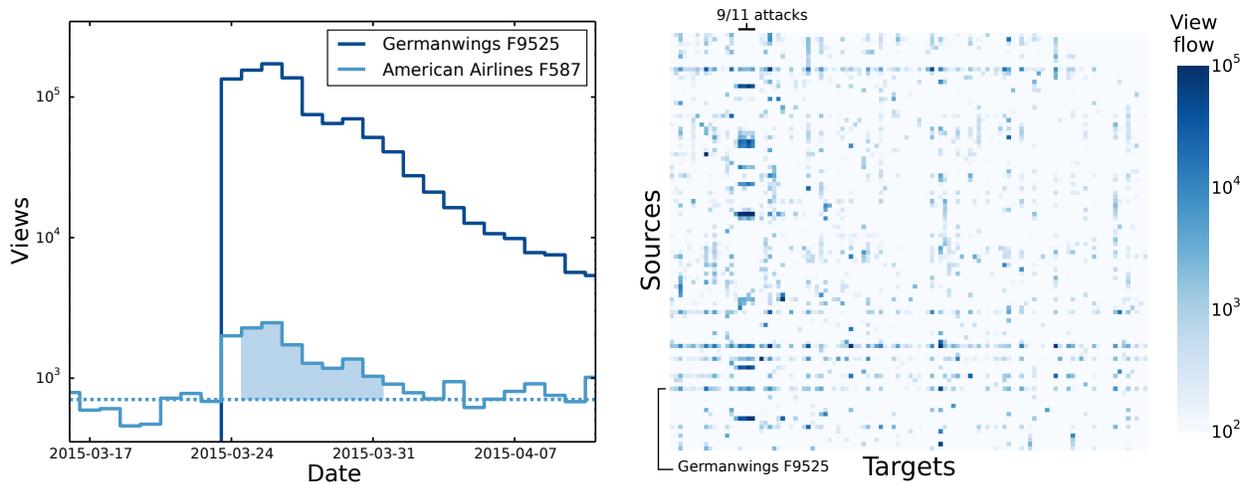}
\caption{\textbf{View flow.} Left panel: Daily Wikipedia article view count on a logarithmic scale for the Wikipedia articles representing Germanwings Flight 9525 (source) and American Airlines Flight 587 (target). 
The colored area measures the increase in views relative to the daily average of the previous year (dashed line). Right panel: View flow from 98 sources (2008-2016) to all 112 target events from the period 2000-2008. The color of the pixels shows the strength of the view flow on a logarithmic scale.  Both axes are sorted according to the date of the accident such that going down or going right brings you to more recent events. Some source events, like Germanwings Flight 9525 (see pointer), trigger a lot of target events. We also point to the articles for the 9/11 crashes, which are triggered often and always in unity. }\label{fig:setup}
\end{figure}

As an example, Figure \ref{fig:setup}A shows the flow of attention from the {\it Germanwings Flight 9525} accident to the {\it American Airlines Flight 587} accident represented by the viewership time series of their corresponding Wikipedia articles. The Germanwings accident occurred on March 24, 2015, when the co-pilot deliberately crashed the plane into a  mountain in the Alps, thereby killing 150 people. The American Airlines accident took place in November 2001 and was caused by a pilot error, which resulted in the plane crashing into the Bell Harbor neighborhood outside New York, thereby killing 265 people. We see an increase in the views to the American Airlines Flight 587 article on the day of the Germanwings crash and this lasts several days. It is worth noting that there was no Wikipedia hyperlink between the two articles during this period. The area of the shaded region measures the increase of the views to the target article relative to the average daily views of the previous year (dashed line), called ``prior activity". We refer to this area as the \emph{view flow} and it will be the central variable of interest in our study. The view flow is calculated over the week after the first edit of the source article. In particular, we focus on the first week as then attention is expected to be maximal \cite{Garcia2016}. Note that any area below the dashed line will count negatively, so the view flow can theoretically be negative as well. 

Our data set includes 84,761 pairs of source and target events (see Data and Methods). In Figure \ref{fig:setup}B we show the view flow from the 98 source events (vertical axis) to all 112 target events from the period 2000-2008 (horizontal axis). We notice that some source events trigger a strong view flow on many target events while others have triggering effect on only few or no target events.  

In the following section we analyse the factors involved in the view flow between each pair in our data set.

\subsection*{Triggering Factors}
\label{subsec:triggering} 

\begin{figure}[ht]
	\centering
	\includegraphics[width=0.95\textwidth]{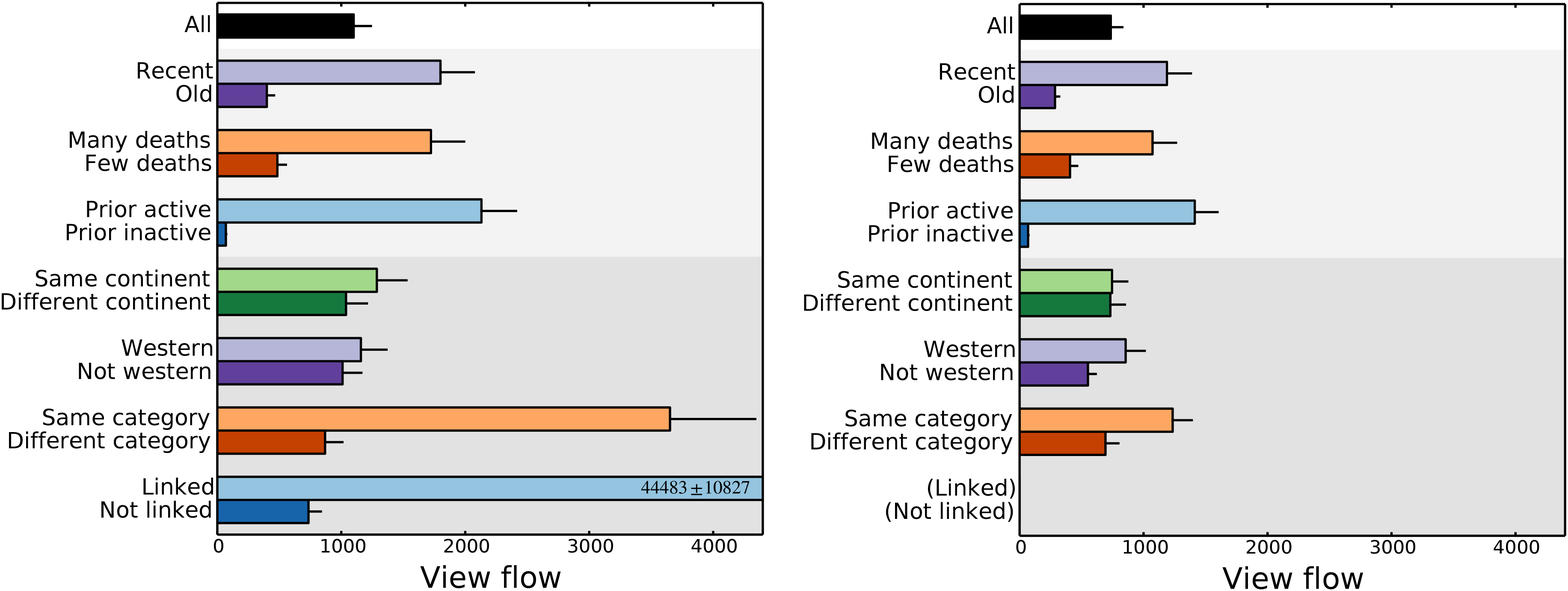}
\caption{\textbf{Triggering factors for view flow.} Left panel: Average view flow among pairs belonging to different groups according to different factors: The black bar labeled ``All" includes all pairs, the bars labeled ``Recent" and ``Old" split the source-target pairs into those that are separated by more or less than 29 years (the median separation between pairs). The bars ``Many deaths" and ``Few deaths" split the pairs according to the number of deaths of the target event (at the median value of 22 deaths). The next two bars split the pairs according to the prior activity of the target article. The bins in dark grey area are based on whether the source and target flights were operated by companies located the same continent, whether the operating company is located in Europe, Australia, or North America (Western), whether the source and target articles belong to the same article categories, and whether there is a direct hyperlink from the source article to the target article. 
Left panel: the same figure as in the left panel, but pairs with a hyperlink from source to target have been removed from the sample.}\label{fig:binary}
\end{figure}

 In this section we limit the analysis to the nine largest sources (8882 source-target-pairs), since the noise is comparable to the signal of the view flow for smaller sources (see Data and Methods). Figure \ref{fig:binary}A shows the average view flow for different groups of source-target article pairs. As expected, we find that target articles about recent events are triggered much more than those about older events. We find that the number of deaths in the target event has an impact: events with more casualties are more likely to be triggered. We also find that the prior viewership of the target articles has a very large impact on the flow of views.  

We do not find a significant effect from the geographical proximity of the two events in the pair. We also find very little impact from the location of the operating company of the target flight. We further check the effect from when the target and source article appear in a common Wikipedia category, as an indication of similarity (see Data and Methods). We find that this has a very large impact on the view flow. Finally, we check if there has been a link from the source article to the target article during any of the seven days under study. We observe that a direct hyperlink has a huge impact on the viewership flow. However, by removing all linked pairs (74 pairs) and performing the same analysis, we get the same qualitative findings (see Figure \ref{fig:binary}B). The average view flow only drops by 33\%, thereby showing that links are not the main driving force responsible for view flow.

\begin{figure}[ht]
	\includegraphics[width=0.99\textwidth]{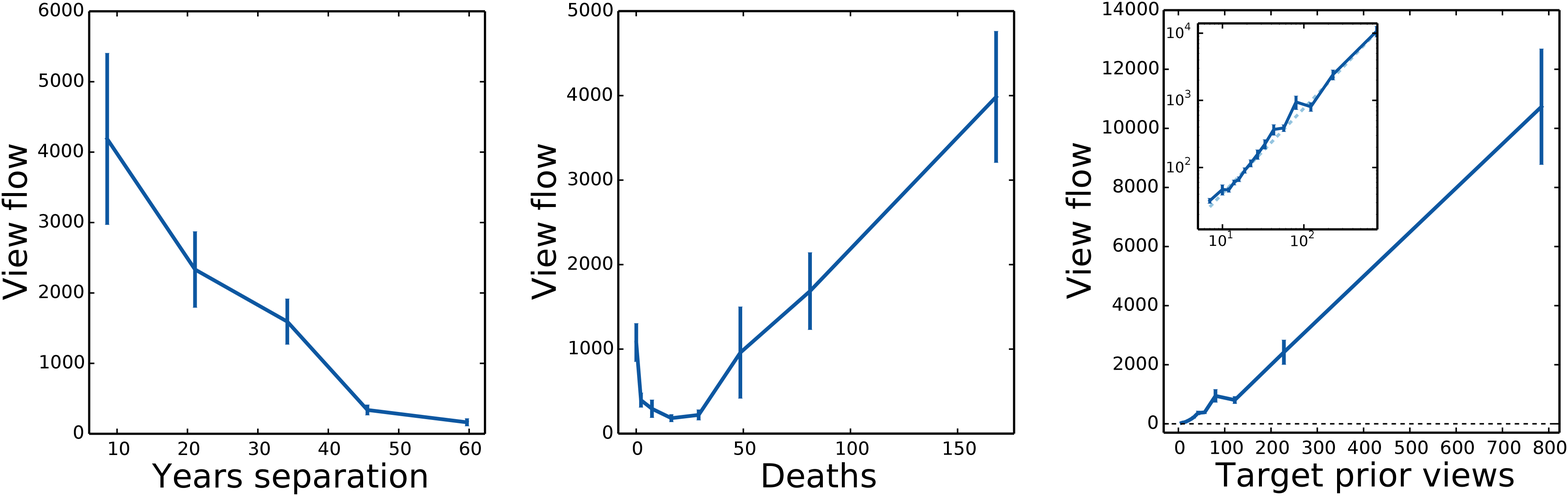}
\caption{\textbf{Detailed analysis of triggering factors for view flow.} Left panel: average view flow against the separation in years between source and target event. Center panel: average view flow against the number of deaths involved in the target event. Right panel: Average daily views of the target article during the year prior to the source event. The insert is in log-log scale and shows a power law fit (dashed) to the curve, which yields an exponent of 1.27. 
}\label{fig:continuous}
\end{figure}

Up to here we analysed the view flow considering all variables as binary, but we can get a better resolution on year separation, deaths, and prior activity. In Figure~\ref{fig:continuous}A, we show the view flow as a function of the separation in years between the source and the target event. The error bars (estimated using bootstrapping) are rather big, but it is clear that we have a strong drop over the first 45 years. In Figure~\ref{fig:continuous}B, we show the view flow as a function of the deaths involved in the target event. As expected, it grows for large number of deaths, but surprisingly, there is a greater flow of views to target articles about events with no deaths compared to those  with $\sim$ 20 deaths. In fact, the average view flow drops from $1076 \pm 227$ to $183 \pm 44$. One possible explanation is that events with zero deaths are reported in Wikipedia because they are remarkable in some other way. Hijackings are a major contributor, but there are other examples such as the 1940 Brocklesby mid-air collision, where two planes collided mid-air and got locked together, but still managed to land safely. In Figure~\ref{fig:continuous}C, we show the view flow as a function of the prior activity of the target article, again measured one year prior to the source event. 
In the insert we show the same graph in log-log scale along with a power law fit. We see that the trend nicely follows the fitted power law with an exponent of $1.27\pm0.04$, showing a super-linear behavior. 

While the source articles, combined, received 6.3 million views during their respective first weeks, we estimate the combined view flow to all the target articles to be 9.8 million. The ratio between the two is $1.56 \pm 0.27$, thereby indicating that the flow of attention is on average greater than the attention received by the main event itself. If we remove all linked pairs, then we are still left with a ratio of $1.03 \pm 0.28$. These results tell us that view flow is no minor player in attention dynamics, but rather a driving force. 

\subsection*{Modeling Remembering}
\label{subsec:modeling remembering}

In the previous section we showed that the online views different topics receive are strongly coupled. Therefore, one cannot describe the attention to a topic as an isolated phenomenon. We will now model the coupling between the source and target and thereby show that a big fraction of the target views may be explained from the source views alone. More formally, we aim to predict the views of a target article $y$ based on the views of a source article $x$ and a number of factors that couple the two. The goal is to maximize the coefficient of determination in predicting $y$. 
We introduce a model with three terms
\begin{equation}
y = y_\textrm{offset} + y_\textrm{link} + y_\textrm{triggered}.
\end{equation}
The first term, $y_\textrm{offset}$, comes from the fact that some target articles receive more attention than others on average. We model this as $y_\textrm{offset} = a_\textrm{history} \cdot y_\textrm{history}$, where $y_\textrm{history}$ is the average weekly views for the previous year. The coefficient therefore takes a value $a_\textrm{history} \sim 1$. With this alone, we are able to explain $15 \pm 4\%$ of the variance among the views of the target articles. Error bars are estimated using bootstrapping. We then include view flow mediated by links ($ y_\textrm{link}$). In order to do this, we estimate the number of views to the source article that are exposed to a link to the target article and call this variable ``exposure to target link" represented as $x_\mathrm{link}$ (see Data and Methods). We then model link flow as $y_\textrm{link} = a_\textrm{link} \cdot x_\textrm{link}$, which in combination with $y_\textrm{offset}$ allows to explain $24 \pm 5\%$ of the variance in views among the target articles. 

The final term in the model, $y_\textrm{triggered}$, represents triggering of memory. Three conditions must be met for a source event to trigger the memory of a target event. First of all, one must hear about the source event and secondly, one must already have the target event stored in long term memory. Finally, the coupling between the two events needs to be sufficiently strong to trigger the memory. We expect the number of people who hear about the source event to be proportional to the number of views to the source article, which we name $x$. Likewise, we expect the number of people who have the target event stored in memory to be proportional to the prior average views of the corresponding target article, $y_\textrm{history}$. Finally, there is the coupling between the two events, $\alpha$, which is the probability that hearing about the source event will trigger the memory of the target event. The first order approximation of the triggered views then can be written as
\begin{equation}
y_\textrm{triggered} = x \times y_\textrm{history} \times \alpha.
\end{equation}\label{eq:model trigger}
For simplicity, we model the coupling using a linear combination of the remaining variables (indexed as $z_i$)

\begin{equation}
\alpha = \sum_i a_i z_i + a_\textrm{0}.
\end{equation}
Here we have not included geographical variables, which proved to be negligible in the above analysis. Instead, we have used information regarding years separation, deaths of the target event, shared Wikipedia category (0 or 1), and target article link (0 or 1). By including the triggering term in the model we increase the explained variance from $24 \pm 5\%$ to $36 \pm 9\%$ (see Data and Methods for parameter values). Note that this is not merely the result of adding 5 more parameters to the model. If we introduce the coupling as an independent term in the model without scaling by $x$ and $y_\textrm{history}$, then we obtain a mere 0.6\% increase in explained variance. Also, if we randomly shuffle around the $y$ values and perform the fit afterwards, then our model captures none of the random variance (a drop from $36\%$ to $0.0\%$), i.e. negligible over-fitting is present. We conclude that our model of source-target flow is capturing some of the true underlying dynamics.


\section*{Discussion}

We introduced ``view flow'', or the attention to an old topic induced by a new topic, as a quantitative measure of remembering. We then used this measure to study the factors of remembering for the case of aircraft accidents and incidents, using data from Wikipedia. In particular, we studied how time, similarity, geography, previous attention, and links  impact the view flow from a source event to a target event. We found that the memory of an aircraft incident effectively lasts around 45 years. This time scale might correspond to people who were adults at the time of the accident not using Wikipedia, dying, or simply forgetting about the accident, such that only written records are left in the end. Incidents with either many (50+) or no deaths are remembered the most on Wikipedia. The last result may be explained by a bias in Wikipedia, which tends to keep records of ``no death'' incidents only if they are remarkable in some other way. 

Generally, we do not find that geographical similarity has any significant impact on remembering of aircraft incidents, even though the level of attention paid to individual incidents is considerably driven by location \cite{Garcia2016}. Links were found to greatly increase the view flow between source and target, but since they are only present for a small fraction of source-target pairs, they cannot explain the majority of the observed view flow. Of more general importance is the previous attention of the target article, which has a super-linear effect on the view flow. This shows that regardless of the strength of the coupling between events, some past events are consistently more memorable. 

The view flow is especially strong when the source and target are similar in some way, as measured by a shared Wikipedia category. Overall, we find that a source event induces a combined view flow, which on average is $\sim 150\%$ of the views given to the source event itself. This tells us that view flow is a major force that should not be ignored. To our knowledge, no current models of online attention and spreading of ideas include coupling between signals. The typical approach in previous studies is to make predictions for the popularity of a topic based on the recent history of that topic alone \cite{pinto2013using,tsur2012s,mollgaard2015emergent, matsubara2012rise}. Future studies should not model attention to a topic using self-correlations only but should also account for cross-correlations to other topics. Concepts, ideas, videos, and so on are not stand-alone objects, but instead form a large network with attention flowing from one to another.

We made a first attempt to model remembering. We proposed to model remembering with a product between the current attention to the source event, the previous attention to the target event, and a coupling between the two. The rationale behind this model is threefold. To trigger the memory of the target event one must hear about the source event, the target event must already be stored in long term memory, and the coupling between the two events must be large enough to trigger the memory. Our model allowed us to explain 36\% of the variance in views among the Wikipedia articles about target events. Note that no information regarding the internal dynamics of the target article views was used to produce this result. A big limitation to our model is the linear expression for the coupling between articles, which, if improved, might allow much more variance to be explained. Furthermore, we do not account for any spreading processes induced by the triggering of memory. Such processes might be responsible for the super-linear relationship observed in Fig. \ref{fig:continuous}C.

Summing up, we argue that the flow of attention between different events and concepts is mediated by memory, or more generally, associativity. We find that source events generate a flow of attention to previous events, which is even greater than the attention given to the source itself. A first model to explain remembering in the case of airline crashes has been provided. The theoretical framework and the mathematical formulation can be easily generalized to explain collective online memory in a broader context.


\section*{Methods}
 
\subsection*{Data Collection}

We collected data from Wikipedia using two main sources: the MediaWiki API and Wikidata.\footnote{Using \url{https://cran.r-project.org/web/packages/WikidataR/index.html}.} The MediaWiki API is a web service that provides access to wiki features, data, and meta-data of articles such as links and categories. Wikidata, on the other hand,  is a Wikipedia partner project that aims to store structured data from other Wikimedia projects, including Wikipedia, and fix inconsistencies across different editions~\cite{Lehmann2015}. Examples of such structured data include the date or geographical coordinates of an event.

We focused on a set of articles in English Wikipedia belonging to the categories \emph{Aviation accidents and incidents by country}  and \emph{Aviation accidents and incidents by year} and their subcategories. These categories 
cover all airline accidents and incidents in different countries and throughout history available in English Wikipedia.
 Using the MediaWiki API, we obtained 1606 articles from which 1496 are specifically  about aircraft crashes or incidents
 (we discard articles of biographies, airport attacks, etc). 
 Furthermore, we extracted editorial information for the articles in the data set: the date when the article was created, alternative names for the article through time, and the article links and categories. We collected the links from the page history for the seven days after the first edit of the article and for each link, we calculated the fraction of the day that it remained in the article.  For the 1496 articles, we systematically collected structured data from Wikidata: the date of the event, geographical coordinates of where the event occurred, number of deaths, and the continent of the aircraft company. Unfortunately, Wikidata did not have complete information for all articles. To remedy this, we obtained the missing data by automatically crawling Wikipedia infoboxes\footnote{Using \url{https://cran.r-project.org/web/packages/WikipediR/index.html}.} or manually checking the information in the articles.

 Finally, we extracted the daily traffic to the articles between 01/01/2008 and 10/04/2016 from the Wikipedia pageview dumps\footnote{\url{https://dumps.wikimedia.org/other/pagecounts-raw/}} through an available interface\footnote{\url{http://stats.grok.se}}. There is no data available before this interval. We used the alternative title names of each article to merge all traffic statistics to the current title.

\subsection*{Sampling}

 The source articles were created in Wikipedia within the period 01/01/2008 -- 10/04/2016. These articles have viewership data available from the moment they were created to the last day of the period. To capture the immediate attention to a source event right after its occurrence, we choose the corresponding  source articles created up to one day after the source event. Furthermore, we remove all source events proximate to a source event with higher attention in the following way: (1) we sort all the source articles by their total number of views during the first week in Wikipedia, (2) starting from the article with the most views, we remove all source articles that were created within a 10 day range and (3) we continue with the next article with highest views and repeat step (2) and so on. In the end, this leaves us with 98 source events.

  We then pair each one of the 98 source events with target events from our entire dataset
  such that each of the target events ocurred at least two years prior to the source event. This assures that the views of the target articles have enough time to stabilize and that we calculate a representative average view rate on the second year, when the views are expected to have already stabilized. This approach leaves us with 84,761 source-target pairs.
  
\begin{figure}[ht]
	\includegraphics[width=0.5\textwidth]{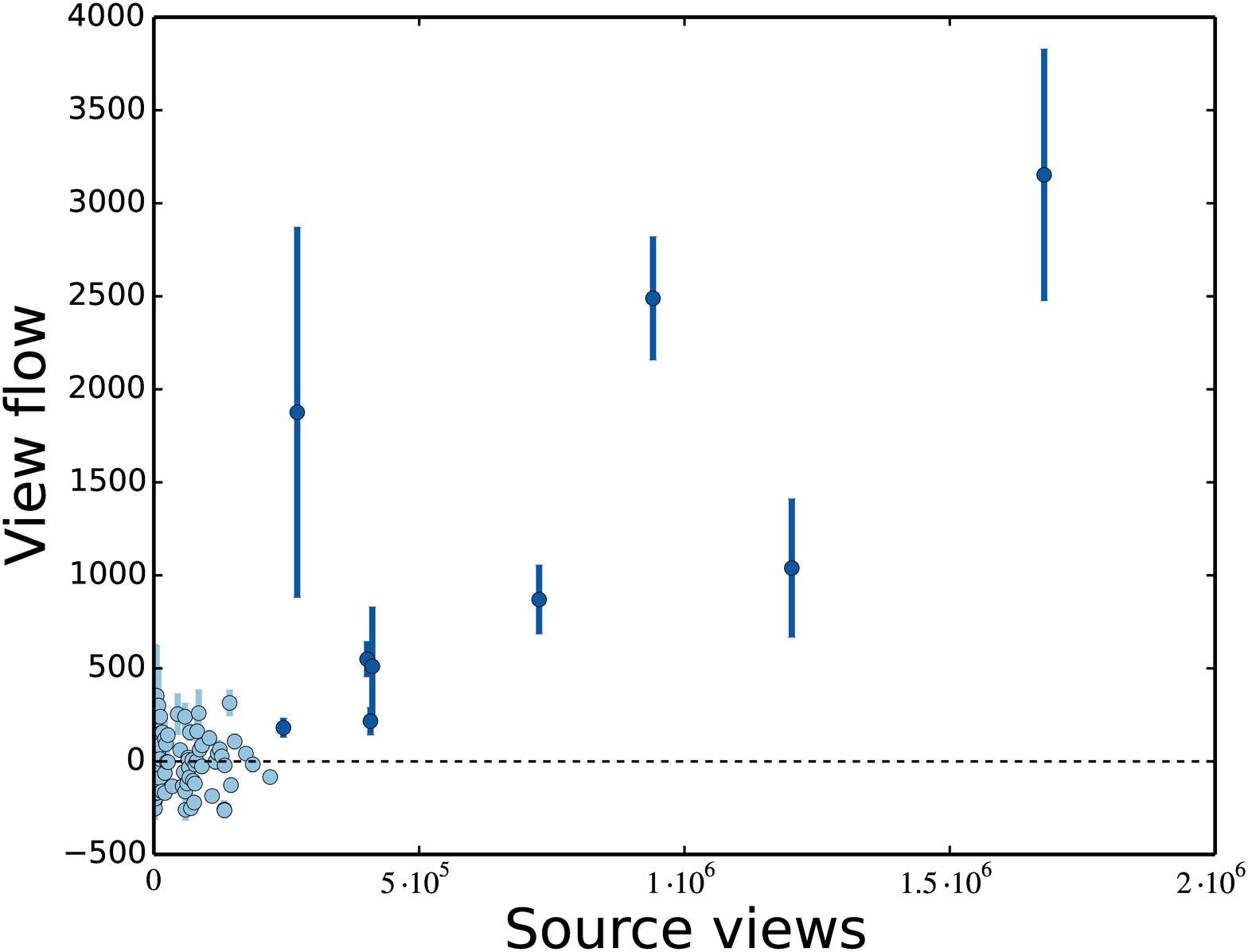}
\caption{\textbf{Average view flow vs. source views.} We show the average view flow from each source against the views of that source during the same period. The dark dots represent the nine largest source events, which were used in section 
\emph{Triggering Factors}: Malaysia Airlines flight 370, Malaysia Airlines flight 17, Air France flight 447, Germanwings flight 9525, 2010 Polish Air Force Tu-154 crash, Indonesia AirAsia flight 8501, Asiana Airlines flight 214, 2011 Lokomotiv Yaroslavl air disaster, and Metrojet flight 9268. The smaller source events (light blue) have not been included because the noise dominates the signal.
}\label{fig:noise}
\end{figure}

In section 
\emph{Triggering Factors} we restricted to the nine largest sources with the argument that the noise of the view flow is comparable to the signal for smaller sources. We illustrate this in Fig. \ref{fig:noise}, which shows the average view flow from any target to its sources. The tenth largest source is the first one where the average is negative, thereby indicating the onset of noise domination.

\subsection*{Category Similarity}

 Categories in Wikipedia form a pseudo-hierarchical structure and their function is to group other regular Wikipedia articles to a common subject \cite{Kittur2009}. In general, categories are socially annotated, and editors can classify an article into a category simply by appending one to it. The categories appended to a Wikipedia article are generally found at the bottom of it. In this project, we consider the common categories among target and source Wikipedia articles as a similarity feature. 

\subsection*{Hyperlinks}

In the context of this project, hyperlinks are internal links in Wikipedia linking a page to another page within English Wikipedia. These blue coloured hyperlinks are an essential feature in Wikipedia since an article can often only be understood in the context of related articles, and internal links make it easy to explore this context \cite{west2015mining}. In this project, we predict the views of the source article flowing to the target article due to an internal link in the source article. To do that, we use ``exposure to target link," $x_{link}$, as an independent input variable for predicting the views of the target article. The variable is calculated using the revision histories of the source articles, which allow us to track the fraction of a given day with an internal link to the target article.  We then construct $x_{link}$ by multiplying this fraction with the number of views of the source article in that day. In the prediction model, we add the resulting number of views for all the days considered in the prediction which in this case is 7 days after the source article was created.
  
\subsection*{Parameter values}

In table \ref{tab:parameters} we show the fitted parameter values with error bars estimated from 10000 bootstrapping samples. The $a_\textrm{linked}$ parameter is part of the coupling constant and should not be confused with $a_\textrm{link}$, which is in the $y_\textrm{link}$ term.

\begin{table}[!htbp]
    \centering
    \begin{tabular}{| l | l | l | l |}
    \hline
    $a_\textrm{history}$ & $a_\textrm{link}$ & $a_\textrm{deaths}$ & $a_\textrm{years}$ \\ \hline
    $0.84 \pm 0.04$ & $0.04 \pm 0.03$ & $2.5\cdot10^{-9} \pm 2.8\cdot10^{-9}$ & $-2.7\cdot10^{-8} \pm 2.1\cdot10^{-8}$ \\ \hline
    $a_\textrm{category}$ & $a_\textrm{linked}$ & $a_\textrm{0}$ \\ \cline{1-3}
    $-1.1\cdot10^{-7} \pm 1.9\cdot10^{-6}$ & $9.7\cdot10^{-6} \pm 5.4\cdot10^{-6}$ & $2.0\cdot10^{-6} \pm 1.0\cdot10^{-6}$ \\ 
    \cline{1-3}
    \end{tabular}
    \caption{Least square fit of the parameters in the model to the data. Error bars are estimated using bootstrapping.}
    \label{tab:parameters}
\end{table}

\section*{Acknowledgements} 
This research is part of the project \emph{Collective Memory in the Digital Age: Understanding Forgetting on the Internet} funded by Google.

\section*{Author contributions statement}
R.G-G. collected and analysed the data, participated in the design of the study, and drafted the manuscript;  A.M. analysed the data, participated in the design of the study, and drafted the manuscript; M.T. participated in the design of the study and helped draft the manuscript; T.Y. designed and coordinated the study, and helped draft the manuscript. All authors gave final approval for publication.

\bibliography{main}  

\end{document}